# Relative importance of social synergy, assortation and networks in the evolution of social cooperation.

Claudia Montoreano and Klaus Jaffe
Universidad Simón Bolívar
Caracas, Venezuela

**Abstract:** We compare the likelihood of different socially relevant features to allow the evolutionary emergence and maintenance of cooperation in a generalized variant of the iterated Prisoners Dilemma game. Results show that the average costs/benefit balance of cooperation is the primary constraint for its establishment and maintenance. Behavior increasing inclusive fitness such as assortation, homophily, kin-selection and tagging of individuals, is second in importance. Networks characteristics were the least important in favoring the establishment and maintenance of cooperation, despite being the most popular in recent research on the subject. Results suggest that inclusive fitness theory with its expansions to include assortative and economic considerations is more general, powerful and relevant in analyzing social phenomena than kin selection theory with its emphasis on genetic relatedness. Merging economics with evolutionary theory will be necessary to reveal more about the nature of social dynamics.

**Introduction**

Cooperation is important in a number of settings, including, behavioral interactions, biological evolution, sociobiology, cultural dynamics, and collective intelligence; yet the features allowing it succeed are not well known. Inclusive fitness theory has been the most successful theory so far in explaining the emergence and maintenance of cooperation in biological systems, but even the father of inclusive fitness theory, W.D. Hamilton, recognized that several different mechanisms are needed to explain the prevalence of social cooperation among extant species (Hamilton 1996). In the search for mechanisms allowing social evolution, the main reason for social behavior, its positive effect on individual fitness, has been neglected. Michener (1969) for example, demonstrated the existence of several different evolutionary routes leading to sophisticated societies that benefited all or most of its members. That is, social cooperation might trigger synergies that increase economic and other benefits to social individuals favoring its evolution (Queller 1992, 2011, Jaffe 2001, 2002, 2010).

Yet the emphasis on genetic aspects of cooperation has sidelined more economic considerations.

That is, the modern emphasis on kin selection rather than on a more general inclusive fitness theory has let to focus on the dynamics of the interactions favoring social cooperation rather than the economic principles that underlie the stability of societies. For example, features of the network of interactions that influence the evolution of cooperation are deemed to be important (Novak 2006). Yet inclusive fitness theory (Hamilton 1964) extended to include assortation (Price 1971) and social synergy, non-additive benefits or positive feedback of social behavior on individual fitness (Queller 1992), has not been pursued with the same enthusiasm. In addition, the features that influence the dynamics of cooperation have been studied using different theoretical frameworks and different virtual setting of games with different specific assumptions. Our aim here is to compare quantitatively, using exactly the same border conditions in the widely used iterated prisoners dilemma game, the relative advantage of the main aspects related to the extended inclusive fitness theory as to its influence on the evolution of cooperation.

Among the different simulation setups used to study the evolution of cooperation, the simplest and most used is based on the Prisoners Dilemma (Axelrod & Hamilton, 1981; Nowak & May, 1993; Riolo et al, 2001; Nowak, 2006; Seinen & Schram, 2006). The prisoner dilemma assumes that synergies are achieved if two agents cooperate, providing benefits to both cooperators; and that there is a relative overall loss and/or an additional cost to the cooperator if one agent does not cooperate. Here we simulate an expanded iterated prisoner's dilemma with a range of costs and benefits to the cooperator that reflect more closely known real situations. In this model, different features affecting the evolution of cooperation can be represented as follows:

- Pay out matrices, punishment, benefits, economic synergies triggered by cooperation, and costs of cooperation, will all affect the relative cost of cooperation to that of not cooperating. The cost/benefit ratio of cooperation might represent all these features. Cost/benefit ratios have shown to be important for the evolution of cooperation in different settings (Nowak & Sigmund, 1998; 2005; Jaffe, 2002: Nowak, 2006; Baranski et al, 2006; Ohtsuki et al, 2006; Jaffe & Zaballa, 2010 )
- Tags that allow cooperators to discriminate agents simulate assortation, as they regulate the type of agents that will interact cooperatively. These tags have been specifically developed for simulations of the prisoner's dilemma as analogy for assortation. The general concept "assortation" includes more specific concepts such as kin selection and assortative mating (Hamilton 1975, Price 1971, Jaffe 2001), and assortative matching or homophily (Riolo et al, 2001; Kim, 2010 and more references discussed below).
- Different types of networks have different effect on the evolutionary dynamics of

cooperation (Zimmerman et al, 2000; Kuperman & Risau-Gusman, 2012; Ohtsuki et al 2006; Kim 2010). The networks with the strongest effect on the likelihood and speed of diffusion of cooperative behavior are random networks and small world networks, in contrast to regular reticular networks which have the weakest effect on this dynamics (Martinez & Jaffe 2012). We thus choose to simulate simple reticular networks and random networks to cover the extremes of this range. The average connectivity K of the network provides for the number of direct neighbors each agent will have. Network interactions can be simulated at various levels, i.e. only with the next neighbor, with expanded neighborhoods, etc. (Ohtsuki et al, 2006, Zhang et al, 2012), increasing the complexity of viscosity of the network. Here we studied both levels of complexity.

**Methods**

We implemented the expanded iterated prisoner's dilemma game published by Riolo et al (2001). Each simulation had 10 000 randomly selected agents placed randomly on the nodes of the network, and in each iteration, each agent decided to cooperated or not cooperate with its neighbors on the network, paying the cost c if the agent engaged in cooperation and always receiving the benefit (b = 10) if at the receiving end of a cooperative interaction. In order to produce different benefit/cost ratios of cooperation, we varied the cost to cooperation from 1 to 10. The simulation consisted of initializing agents and networks and then letting each agent interacts with is neighbors. After all agents have participated in all pairings in a generation, the fitness score of each agent was calculated by adding costs and benefits of all tournaments during this generation. Agents then reproduced on the basis of their fitness score relative to others. That is, we compared each agent with its immediate neighbors, and substituted in the next round, the characteristics of the lower fitness agent with the ones having the higher score. This algorithm produced strong selection, making the outcome of the simulation more dependent on the initial condition where agents and their characteristics were established at random.

When simulating assortation, we followed Riolo et al (2001) to model agents with tags. Each agent was assigned a random Tag ($t \in [0,1]$) and a Tolerance ( ), so that an agent would cooperate with a neighbor only if
$$t_n \in (t_a - T_a, t_a + T_a)$$
where $t_n$= neighbor's tag, $t_a$= agent's tag, $T_a$= agent's tolerance. (Non-cooperators T=0)

The networks used where a Regular-grid Graph where each agent could have 1, 4 or 8 neighbors (k=1, k=4, k=8) and a Random Erdös-Rényi Graph (Erdös and Rényi, 1959) with an average of 4 neighbors (k=4). For the Regular Graph with k=1, each agent had a neighborhood of 4 and selected randomly a single individual to interact in each iteration, and for k=4 and k=8 we used the neighborhoods of Von Neumann and Moore respectively.

The effect of extended neighborhood was studied by allowing agents to have contact with their neighbors and with their neighbor's neighbors, with a given probability of interaction: each agent had its regular neighborhood, and with a probability of 0.8 it would interact with someone in the extended neighborhood.

Each simulation run was performed with a fixed set of characteristics for agents (cooperator or non-cooperator, with or without tags) and networks (regular grid or random networks, with or without expanded neighborhoods, with different degrees of connectivity K), producing 12 different combinations as reported in Table 1. For each case we ran simulations for costs from 1 to 10, for initial fraction of cooperators from 1% to 100%, and with 20 replicated for each combination of parameters, with a maximum of 500 iterations for each simulation (31).

**Results**

Different benefit-cost ratios (b/c) were produced by varying c from 1 to 10 and leaving b=10 constant. Simulations with values of c above 10 never produced a majority of cooperators and values of c=b=10 (b/c=1) produced random distributions of cooperators among non-cooperators with neither displacing the other.

Figure 1 shows the temporal changes of the output of a simulation when b/c > 1. Starting with a few cooperators (white spaces), after a single time step, isolated cooperators go extinct as they lose in the fitness game against non-cooperators (black spaces) when they have no cooperative neighbors. Then groups of neighboring cooperators out-breed non-cooperators increasing the total population of cooperators, until, depending on the spatial structure, they displace all non-cooperators or reduce their populations to small redoubts in the virtual space. The simulation shows a final equilibrium situation with an approximately constant final average frequency of cooperators which is slightly below 1.

**Figure1.** Change in the cooperator's fraction in a simulation. Initial cooperator's fraction 0.13, cost to cooperation 1, agents without tags, without learning and without extended neighborhood in a regular grid graph.

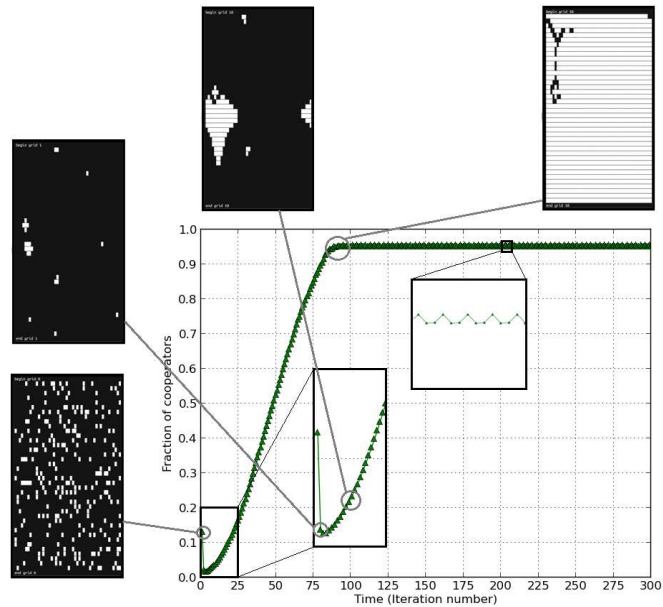

**Figure 2.** Final average percentage of cooperators obtained in simulations with different initial percentage of cooperators, at a fixed cost C.

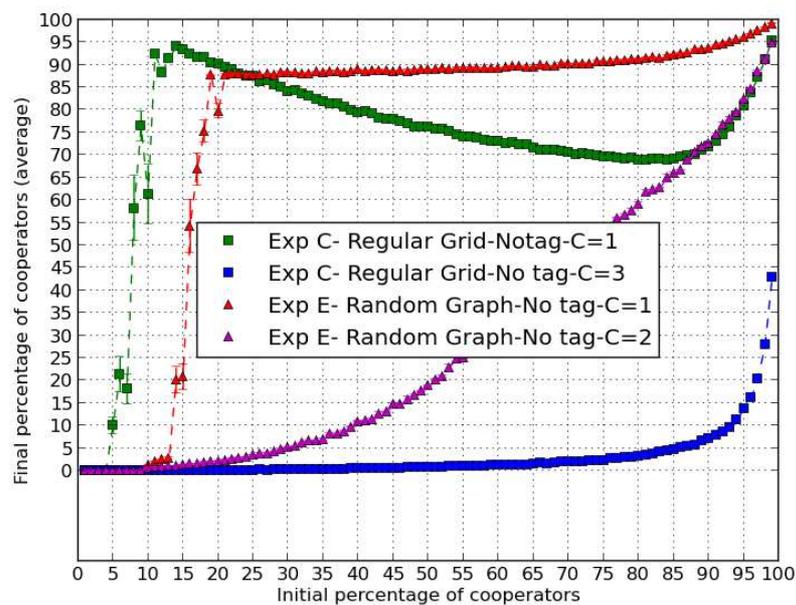

In Figure 2 we show the average percentage of cooperators at equilibrium at the end of simulations, when simulations are started with different initial percentages of cooperators. The results show that when the initial number of cooperators is too small they are more likely to go extinct, reducing the average equilibrium percentage of cooperators at the end of the simulations. When simulations start with higher percentages of cooperating agents, the chances that they out-breed non-cooperators are higher. Some interesting exceptions were revealed. For example, simulations using Regular-G and C= 1 show that initial percentages of cooperators of about 80 produce lower average final percentages of cooperators tan initial percentages of cooperators around 20. This effect is due to the specific dynamics induced by the network characteristics which favors or hinders the establishment of clusters of cooperators. In this specific example, isolated groups of non-cooperators are able to infiltrate the mostly cooperative clusters. In consequence, cooperators are more likely to appear when the initial percentage of cooperators is not too high; although at initial percentages of cooperators above 80, cooperation is stable even outside isolated groups of cooperators. In some settings (C=3 for example), only very high initial number of cooperators allows them avoid being totally displaced by non-cooperators.

**Figure 3.** Phase diagram: Initial percentage of cooperators for each cost for which each final average percentage of cooperators is at least 50%

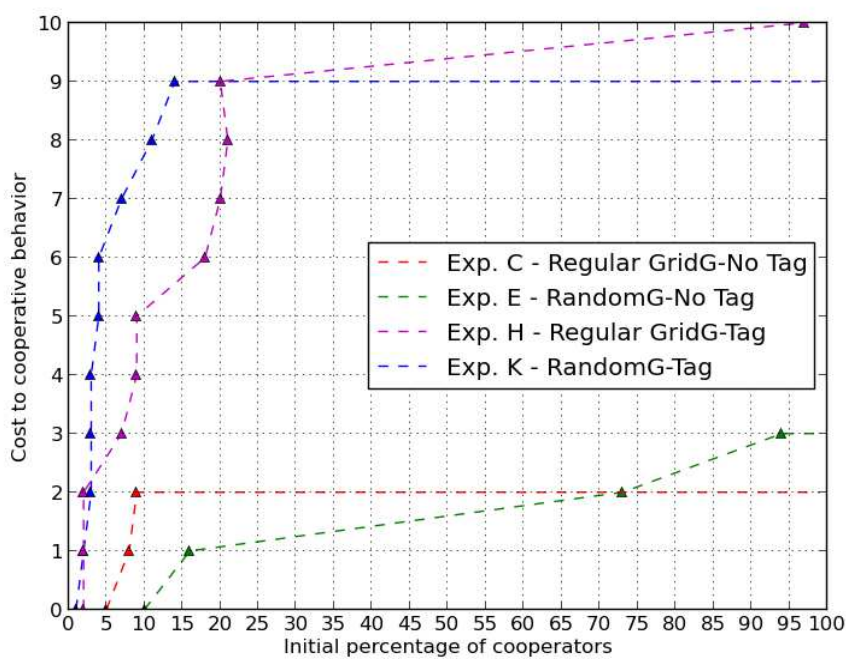

Figure 3 shows an example of a phase diagram, plotting the initial number of cooperators required in a simulation to eventually out-breed non cooperators, at different costs (c) of the cooperative act.

7Different simulation scenarios allow cooperators to out-breed non-cooperators at different maximal costs of cooperation and at different initial percentages of cooperators in the simulation. In this particular plot, we observe that simulations using agents with tags produce stable populations of cooperators even if the costs of cooperation are as high as c=9. Agents without tags stabilize populations of cooperators only if c < 4.

**Table 1:** Summary of the results for every setting considering maximum affordable cost for cooperators to invade a population and to stay in it (Initial Fraction of cooperators 0.25 and 0.50; final fraction of cooperators >= 50%)

| Graph | Exp | K | Expanded Neighbors | Tag | Cost IF=.25 | Cost IF = .50 |
|---|---|---|---|---|---|---|
| Regular-grid G | A | 1 | | | 0 | 0 |
| Regular-grid G | B | 8 | | | 2 | 2 |
| Regular-grid G | C | 4 | | | 2 | 2 |
| Regular-grid G | D | 4 | Yes | | 1 | 1 |
| Random G | E | 4 | | | 1 | 2 |
| Random G | F | 4 | Yes | | 1 | 2 |
| Regular-grid G | G | 1 | | Yes | 0 | 9 |
| Regular-grid G | H | 4 | | Yes | 9 | 9 |
| Regular-grid G | I | 8 | | Yes | 9 | 9 |
| Regular-grid G | J | 4 | Yes | Yes | 9 | 9 |
| Random G | K | 4 | | Yes | 9 | 9 |
| Random G | L | 4 | Yes | Yes | 8 | 9 |

Table 1 summarizes our results showing the maximum cost of cooperation that each system can suffer before cooperation collapses. We refer to a collapse of cooperation when the simulation conditions favor the establishment of a majority of non-cooperative agents in the equilibrium population. A higher maximum cost signals a more robust systems regarding the establishment and maintenance of cooperation, as simulations evolve equilibrium populations with a majority of cooperators despite higher costs. The results show that a cost benefit ratio can be always found that favors the establishment of cooperation. Then, the greatest effect in stabilizing the odds for cooperative strategies to evolve in a virtual population of interacting agents is the presence of a tag that allows discriminating between cooperators and non-cooperators. For example, cooperation using random grids without tags and without expanded neighborhood (experiment E) collapsed when cost of cooperation was equal or greater than 2. In contrast, cooperation in the same experiment but using tags (experiment K) collapsed with costs of cooperation of 9 or above. That is, assortation increased the robustness of cooperation approximately 4.5 times (9/2). Assortation, simulated with tags that allowed agents to cooperate more with agents similar to them, favored the establishment of cooperation even at relatively high fitness cost to cooperator.

Expanding cooperative interactions to second neighbors reduced slightly the maximum cost of cooperation that supported stable populations of cooperative agents Compare experiment C with D, E with F, J with L or M with N). That is, cooperation is more stable when cooperators can avoid non-cooperators.

The connectivity of the grid as expressed by K seems to be important at low values of K as shown by comparing experiment A *vs*. C. Increasing K further does not improve the fate of cooperators (compare experiment B with C). The grid structure had little impact on the stability of populations of cooperators (compare experiments D *vs*. F and H *vs*. K). Regular grids were a little more robust, supporting higher costs of cooperation than random ones (compare experiments C *vs*. E and J *vs*. L with low initial numbers of cooperators). Again, this might be due to the fact that barriers protecting cooperators might stabilize their evolution. More detailed results are presented elsewhere (Montoreano 2012).

**Discussion**

Regarding the factors that are deemed relevant to our understanding of the emergence and maintenance of social cooperation and the stability of societies, there is a clear hierarchy in their importance. Results on Table 1 show that in any of the situations tested, there is an economic condition that allows for the evolution of cooperation. The second largest effect was produced by tagging, suggesting that assortation is indeed very important in aiding social cooperation. These results are congruent with other simulations using completely different games, such as sociodynamica (Jaffe 2002) and others, showing that economic synergy and assortation favor the establishment and maintenance of social cooperation.

The average likelihood to receive short and long term economic benefits favors the establishment and maintenance of cooperation. Simple short term cost-benefit analyses might overlook this likelihood. The punishment of non-cooperators, which has been found to strongly favor evolution of cooperation (Nowak & Sigmund, 1998; 2005; Nowak, 2006; Baranski et al, 2006; Jaffe & Zaballa, 2010) is equivalent in our model to a high benefit cost ratio of cooperation, as the consequence of punishment is to increase the cost to non-cooperators.

Tags; which in this simple simulation scenario can be used as analogies for assortation which includes homophily, green-beards, kin-structure, assortative matching, viscosity in the dispersion of kin, etc; was the second most important feature allowing the stability of populations of cooperative agents. This supports the use of Price´s extension of the Hardy-Weinberg to include assortation, favoring the concept of inclusive fitness over that of kin-selection as a better representation of reality. Price´s equation (Price 1971) tacks the change in mean phonotype due to selection, but does not account for positive effects on selection elicited by the act of cooperation itself. This social synergy effect has been incorporated in a general equation accounting for social evolution (Queller 1992) and is recognized as of primary importance in the economics and business literature.

The working of assortation in favoring the success of cooperative strategies seem to be associated with the possibility of forming globular clusters, as is the case of some network structures (Kuperman & Risau-Gusman, 2012). Yet assortation is much more powerful in favoring cooperation than the networks studied here. This seems to be related to a more general phenomenon suggesting that behavior is more flexible than structure in adapting to complex adaptive landscapes. This would explain our results. Thus our results suggest focusing future research efforts economic aspects and on the mechanisms allowing the working of assortation, such as the dynamics between evolution of tags and evolution of cooperation, and caution that the popularity of network research might not help explain the most important aspects of the evolution of cooperation.

Networks, very popular in the modern literature about the dynamics of cooperation had been proposed to help the evolution of cooperation (for example, Zimmerman et al, 2000; Hofbauer and Sigmund, 2003, Ohtsuki et al, 2006; Pacheco et al, 2006, Kun and Scheuring, 2009; Kim, 2010; Zhang, 2012). Our results show that the degree of connectivity and the type of network affect the dynamics of cooperation, but much less than the cost/benefit balance, and of mechanisms facilitating assortation. Networks might favor isolation required by groups of cooperators to cope with social parasites.

The main conclusion that can be draws then is that social synergy or economic benefits that derive from social life seem to be the strongest driver in the evolution of cooperation in this game, which has been widely used as an illustrative metaphor of real life among physicists, biologists, political scientists and economists alike. This finding shows that the expanded version of Inclusive Fitness Theory is more relevant than Kin Selection Theory, as it incorporates social synergy in its equations, which Kin Selection with its exclusive emphasis on genetic relatedness does not. Little quantitative empirical research on social synergy has been produced in biology (but see Osborne & Jaffe 1997, Jaffe 2010, Smith et al 2010), though it is recognized as of primary importance in the economics and business literature. Thus, a synergetic interchange of theoretical knowledge between economics and biology looks promising for a novel attempted to deepen our understanding of social dynamics.

**Acknowledgements:** We thank Ivette Carolina Martinez, Marcelo Kuperman and David Queller for helpful comments and sharp criticism of earlier drafts of the manuscript.